# Nonlinear Photoluminescence in Atomically Thin Layered WSe$_2$ Arising from Diffusion-Assisted Exciton–Exciton Annihilation

Shinichiro Mouri,[1] Yuhei Miyauchi,[1,2,3]

Minglin Toh,[4] Weijie Zhao,[5,7] Goki Eda,[5,6,7] and Kazunari Matsuda[1]

[1]*Institute of Advanced Energy, Kyoto University, Uji, Kyoto 611-0011, Japan*

[2]*Japan Science and Technology Agency, PRESTO, 4-1-8 Honcho Kawaguchi, Saitama 332-0012, Japan*

[3]*Graduate School of Science, Nagoya University, Chikusa, Nagoya 464-8601, Japan*

[4]*School of Materials Science and Engineering, Nanyang Technological University, N4.1 Nanyang Avenue, Singapore 639798, Singapore*

[5]*Department of Physics, National University of Singapore, 2 Science Drive 3, Singapore 117542, Singapore*

[6]*Department of Chemistry, National University of Singapore, 3 Science Drive 3, Singapore 117543, Singapore*

[7]*Graphene Research Centre, National University of Singapore, 6 Science Drive 2, Singapore 117546, Singapore*

## Abstract

We studied multi-exciton dynamics in monolayer WSe$_2$ using nonlinear photoluminescence (PL) spectroscopy and Monte Carlo simulations. We observed strong nonlinear saturation behavior of exciton PL with increasing excitation power density, and long-distance exciton diffusion reaching several micrometers. We demonstrated that the diffusion-assisted exciton–exciton annihilation model accounts for the observed nonlinear PL behavior. The long-distance exciton diffusion and subsequent efficient exciton–exciton annihilation process determined the unusual multi-exciton dynamics in atomically thin layered transition metal dichalcogenides.

Many-body effects arising from strong interactions among carriers have been of considerable interest in nanoscience. Spatial confinement and reduced Coulomb screening in low-dimensional nanostructures enhance many-body scattering processes such as carrier multiplication[1–4] and Auger recombination.[5–7] Exciton–exciton annihilation (EEA) is one of the key processes for understanding the dynamics of optically excited states in low-dimensional materials such as carbon nanotubes[4,5,8–12] and nanocrystals.[13,14] However, despite its importance, there is very limited knowledge of EEA in two-dimensional (2D) electronic systems, such as quantum wells,[15] presumably because imperfect confinement and relatively strong dielectric screening in conventional quantum wells composed of compound semiconductors weaken the Coulomb interaction that is responsible for the EEA rate.

Atomically thin layered transition metal dichalcogenides (TMDs), i.e., $MX_2$ (M = Mo, W, and X = S, Se), have been extensively studied as novel 2D semiconductors[16–25] from the viewpoint of fundamental physics and various optoelectronic applications such as phototransistors,[26] light-emitting devices,[27] and solar cells.[28] Moreover, they are a promising platform to study the strong interactions among optically generated carriers because of their strong quantum confinement and reduced Coulomb screening. The binding energy of an optically excited electron–hole pair (exciton) in TMDs reaches >500 meV,[29,30] which is about one order of magnitude larger than that of compound semiconductor quantum wells.[16] Indeed, the interesting features of excitons and exciton-carrier interactions[31–36], including the EEA process[37], have been studied in few-layered TMDs; however, the detailed dynamics and mechanism of the EEA process in TMDs has not been clarified. The elucidation of the dynamics and mechanism of EEA in monolayer

TMDs is a challenging issue for understanding many-body correlations between excitons in extremely confined atomically thin layered 2D systems. Moreover, the EEA process reduces the generated exciton density and therefore affects the performance of optoelectronic devices; thus understanding EEA is essential for improving the performance of such devices.

In this paper, we studied the many-body correlations between excitons in monolayer $WSe_2$ (1L-$WSe_2$) by excitation power dependence of photoluminescence (PL) and time-resolved PL spectroscopy. We determined that excitons in 1L-$WSe_2$ demonstrate long-distance migration reaching ~1.8 μm at room temperature. The strong nonlinear saturation behavior of PL intensity with increasing excitation power density due to the appearance of a rapid exciton PL decay component was observed under very low excitation power regimes. Using computational simulations based on the Monte Carlo method, we demonstrated that diffusion-assisted EEA can account for the rapid exciton decay component and the nonlinear saturation behavior of exciton PL. The long-distance exciton diffusion and subsequent efficient EEA because of the enhanced Coulomb interaction determine the unusual multi-exciton dynamics in atomically thin layered TMDs.

The 1L-$WSe_2$ flake on the quartz substrate was mechanically exfoliated from the single crystal prepared by chemical vapor transport method.[36] The thickness of the $WSe_2$ flake was verified by optical contrast and atomic force microscopy, as shown in Ref. 38. A monochromated light pulse (2.33 eV) from a super-continuum broadband light source (40 MHz, 20 ps pulse duration) was used as an excitation source. All optical measurements were conducted using a home-built micro-PL setup at room temperature.

PL spectra were detected by a monochromator equipped with a liquid $N_2$-cooled CCD camera. Time-resolved PL decay profiles of 1L-WSe$_2$ were recorded using a time-correlated single-photon counting technique, and the time resolution of the setup was estimated from the instrumental response function (IRF) as ~100 ps. PL signals through band-pass filters with a bandwidth of 20 meV were detected using an avalanche photodiode for the time-resolved measurement.

Figure 1(a) shows the PL spectrum of 1L-WSe$_2$ under weak excitation conditions (0.03 μJ/cm$^2$). A strong PL peak was observed at 1.67 eV, which is assigned to the exciton recombination due to the direct band gap transition[38,39]. The small tail of the PL spectrum on the lower-energy side is due to PL from trions or defect-trapped excitons,[25,31,40,41] which suggests a low unintentionally doped carrier density in this sample.

Figure 1(b) shows the PL decay profile of 1L-WSe$_2$ at an excitation power density of 0.03 μJ/cm$^2$. The PL decay curve shows the two decay components. The PL decay curve within the first few nanoseconds simply shows the exponential decay behavior, and the decay time, $\tau$, was evaluated as ~4 ns. We assigned a PL decay time of ~4 ns as the recombination lifetime of excitons in 1L-WSe$_2$. The observed exciton lifetime of ~4 ns in 1L-WSe$_2$ is much longer than that reported in 1L-MoS$_2$ (<100 ps).[33,41] This is physically reasonable, because the PL quantum yield of 1L-WSe$_2$ (>~$10^{-1}$) is much higher than that in 1L-MoS$_2$ (~$10^{-3}$). The observed slow decay component after 4 ns with a decay time of ~12 ns may be due to PL of defect-trapped excitons.[40]

Figure 1(c) shows the PL image obtained by a focused laser under weak excitation conditions (<0.03 μJ/cm$^2$) with the corresponding excitation laser profile in the inset. It was found that the obtained PL image of an exciton was larger than the excitation laser profile. This suggests the spatial diffusion of optically generated excitons in 1L-WSe$_2$. Figure 1(d) shows the cross-sectional profile of the PL image indicated by the line in Figure 1(c). We fitted the cross-sectional profile of the PL image with a Gaussian function derived from the exciton diffusion model (solid red line),[42] in which the laser spot profile as an IRF was considered to estimate the exciton diffusion length $L_{ex}$. From this analysis, $L_{ex}$ was evaluated as a large value of 1.8 (±0.5) μm in 1L-WSe$_2$ at room temperature ($T$ = 300 K). This result suggests that the optically generated excitons move diffusively over a long distance in the order of micrometers during the recombination lifetime. The exciton diffusion coefficient, $D_{ex}$, was derived as ~2.2 (±1.1) cm$^2$/s using the relationship, $L_{ex} = 2\sqrt{D_{ex}\tau_{ex}}$, where $\tau_{ex}$ is the exciton lifetime of ~4 ns under weak excitation conditions. The experimentally obtained $D_{ex}$ value of ~2.2 cm$^2$/s is almost consistent with the estimated value of ~2.0 cm$^2$/s from the Einstein relation $D_{ex} \approx k_B T \Delta / M_{ex}$, where $k_B$ is the Boltzmann constant, $\Delta$ is the homogeneous linewidth of excitons (~40 meV) determined by reflectance measurements, and $M_{ex}$ is the exciton translational mass (~0.68 $m_0$).[43]

Figure 2(a) shows the normalized PL spectra of 1L-WSe$_2$ on a quartz substrate, obtained by increasing the excitation power density, where each PL spectrum is normalized by the corresponding excitation power density. The PL intensity increases with increasing excitation power density; however, the relative intensity of the normalized PL spectrum gradually decreases as shown in Figure 2(a), which suggests strong saturation behavior of the PL spectra as a function of excitation power density. Moreover, the shapes of the PL spectra are unchanged compared with those at weak (0.006 µJ/cm$^2$) and strong (12 µJ/cm$^2$) excitation power densities, as shown in Figure 2(b). This suggests that the strong nonlinear behavior of the PL spectra is due to the exciton dynamics.

Figure 2(c) shows the integrated PL intensity of 1L-WSe$_2$ as a function of excitation power density. The PL intensity simply increases linearly as a function of excitation power density under low excitation conditions (<0.03 µJ/cm$^2$). In contrast, the PL intensity deviates from simple linear dependence and gradually saturates as a function of excitation power density above ~0.06 µJ cm$^{-2}$, which shows the strong nonlinear PL behavior, as shown in Figure 2(a). Such a strong nonlinear PL response is quite different from the previously reported linear response in MoS$_2$[41] and WSe$_2$.[25]

To understand the origin of this strong nonlinear PL behavior, we measured the time-resolved PL decay in 1L-WSe$_2$ at various excitation power densities. Figure 3 shows the change in the PL decay curves with increasing excitation power density. Under low excitation conditions (<0.03 µJ/cm$^2$), the shape of PL decay profile did not change. In contrast, under higher excitation conditions, the rapid PL decay component noticeably appeared with increasing excitation power density. The PL decay curves are composed of the rapid decay and bi-exponential decay components with time constants of ~4 ns and 12 ns, respectively. We calculated the time-integrated PL intensity from the integration of the decay profile over the entire time range at each excitation power density. The time-integrated PL intensity is plotted in Figure 2 (b) (blue squares), and the excitation power dependence of the time-integrated PL intensity is almost coincident with that of the PL intensity obtained from the PL spectrum. Thus, the appearance of the rapid decay component depending on the excitation power density primarily contributes to the strong nonlinear response of exciton PL in 1L-WSe$_2$. Moreover, it was noted that the rapid decay component appears even in very weak excitation region below 0.06 μJ/cm$^2$ in Figure 3, in which the optically generated exciton density at time $t = 0$ ($n_{ex}^0$) estimated from the excitation power density and absorbance of 1L-WSe$_2$ at 2.33 eV (~0.06 μJ/cm$^2$) is of a very low value of <1.2 × 10$^{10}$ cm$^{-2}$. This value corresponds to a very large average distance between optically generated excitons (>~130 nm), which is much larger than the exciton Bohr radius, $a_B$ ~0.6 nm (see Supporting Information 1).

Here, we examine the diffusion-assisted EEA scheme as a possible mechanism for the emergence of the rapid exciton decay component with increasing excitation power density, even under low excitation power densities. Figure 4(a) shows a schematic of diffusion-assisted EEA, i.e., EEA occurs with the probability $P_{EEA}$ and one of the excitons is nonradiatively relaxed to the ground state when two excitons are encountered after the long-distance exciton diffusion. We conducted a computational simulation of multi-exciton decay dynamics with consideration of the diffusion-assisted EEA process based on the Monte Carlo simulation method (see Supporting Information 2). Figure 4(b) shows that the simulated exciton decay curves are proportional to the PL intensity at various initial exciton densities $n_{ex}^0$ obtained by the Monte Carlo method, where the parameters were set as $P_{EEA} = 0.25$, $D_{ex} = 2$ cm$^2$/s, and $\tau_{ex} = 4$ ns. With increasing initial exciton density, the rapid decay component due to EEA becomes dominant, similar to the experimentally observed PL decay curves shown in Figure 3. The simulation results indicate that this simple model can explain the experimentally observed rapid exciton PL decay in 1L-WSe$_2$, even when the optically generated exciton density is small.

Moreover, we conducted calculations with various values of $P_{EEA}$ and determined that $P_{EEA} = 0.25$ was the most suitable value to reproduce the time scale of the rapid decay component for all of the excitation power densities (see Supporting Information 3). The inset of Figure 4 shows the comparison between the half-decay time of the rapid decay component estimated from the simulation (red circles) and that from the experimental result (black squares) in Figure 3. The simulated half-decay time is consistent with the experimentally obtained half-decay time, which supports the validity of this simulation. The EEA rate $\gamma_{EEA}$ was estimated as ~0.35 cm$^2$/s using the relation $P_{EEA} = \gamma_{EEA} dt/A$, where $A$ ($= 4\pi a_B^2$, ~4.5 nm$^2$) is the overlapping area of an exciton and $dt$ (= 32 fs) is the time of the computational step determined by exciton dephasing. Note that the evaluated value of the EEA rate (~0.35 cm$^2$/s) is much larger than that in compound 2D semiconductor quantum wells (~10$^{-3}$ cm$^2$/s).[15] The enhanced Coulomb interaction due to strong quantum confinement within atomically thin layers plays an essential role in an efficient EEA process in 1L-TMDs, because the EEA rate is determined by the direct Coulomb interaction. According to theoretical calculations and recent experimental studies, the binding energy of excitons has been evaluated as >0.5 eV because of the enhanced Coulomb interaction in 1L-TMDs.[29,30,43–45] The value is one order of magnitude larger than that in compound 2D semiconductor quantum wells.[46] Thus, anomalous multi-exciton dynamics of diffusion-assisted EEA occurs because of a very long exciton diffusion length of several micrometers and an enhanced Coulomb interaction in atomically thin layered 2D TMDs.

In summary, we studied exciton–exciton interactions in 1L-WSe$_2$ using PL spectroscopy and Monte Carlo simulations. We determined the strong nonlinear saturation behavior of exciton PL as a function of excitation power density, and long-distance exciton diffusion reaching 1.8 μm in 1L-WSe$_2$. We demonstrated that the diffusion-assisted EEA model accounts for the experimentally observed strong nonlinear PL behavior. The long-distance exciton diffusion and subsequent efficient EEA with an extremely large rate of ~0.35 cm$^2$/s result in the unusual multi-exciton dynamics in atomically thin layered TMDs. These results provide important insights for understanding exciton–exciton interactions in monolayer TMDs and have the potential to enhance the performance of optoelectronic devices such as solar cells and high-efficiency photodetectors based on TMDs.

## ACKNOWLEDGMENTS

The authors acknowledge S. Konabe and S. Okada for fruitful discussions about theoretical part. This study was supported by a Grant-in-Aid for Scientific Research from MEXT of Japan (Nos. 22740195, 25400324, 24681031, 23340085, and 25610074) and by PRESTO from JST. G. E acknowledges Singapore national Research Foundation for funding the research under NRF Research Fellowship (NRF-NRFF2011-02).

**FIGURE CAPTIONS**

**Figure 1.** (a) PL spectrum of 1L-WSe$_2$ excited by 2.33 eV light at a power density of 0.03 μJ/cm$^2$. (b) Time-resolved PL decay profile of 1L-WSe$_2$ excited by 2.33 eV light at 0.03 μJ/cm$^2$. The gray curve shows the instrumental response function (IRF). (c) PL image of 1L-WSe$_2$. The inset shows the laser profile. (d) Cross-sectional profile of excitation laser (IRF) and PL image of 1L-WSe$_2$. The solid red line shows the calculated cross-sectional profile of the PL image with a Gaussian function derived from the exciton diffusion model.

**Figure 2.** (a) PL spectra of 1L-WSe$_2$ with varying excitation power density. PL spectra are normalized by the corresponding excitation power density. (b) PL spectra of 1L-WSe$_2$ normalized by their peak values with excitation power densities of 12 and 0.006 μJ/cm$^2$. (c) PL intensity as a function of excitation power density (black circles). Calculated PL intensity deduced from the integration of the PL decay profile over the whole time range as a function of excitation power density (blue squares).

**Figure 3.** PL decay profiles of 1L-WSe$_2$ with varying excitation power density. The vertical axis for PL intensity is a logarithmic scale.

**Figure 4.** (a) Schematic of the diffusion-assisted exciton–exciton (EEA) annihilation process in an atomically thin layered material. (b) PL decay curves with varying excitation power density obtained from Monte Carlo simulations using EEA probability

$P_{EEA} = 0.25$ and diffusion coefficient $D_{ex} = 2$ cm²/s. The vertical axis for PL intensity is a logarithmic scale. The inset shows the half-decay time estimated from experimental PL decay curves (black squares) and from calculated exciton decay curves (red circles) as a function of estimated exciton density.

# REFERENCES

*E-mail: iguchan@iae.kyoto-u.ac.jp

+E-mail: matsuda@iae.kyoto-u.ac.jp

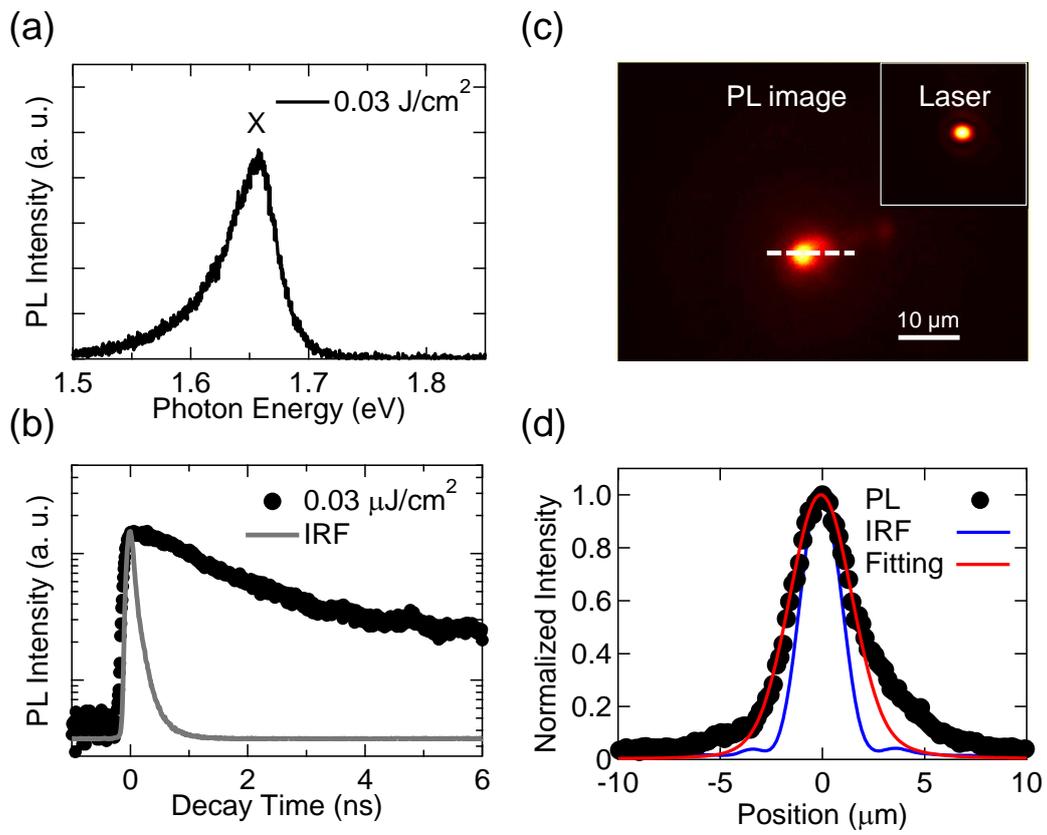

Fig. 1 S. Mouri et al.

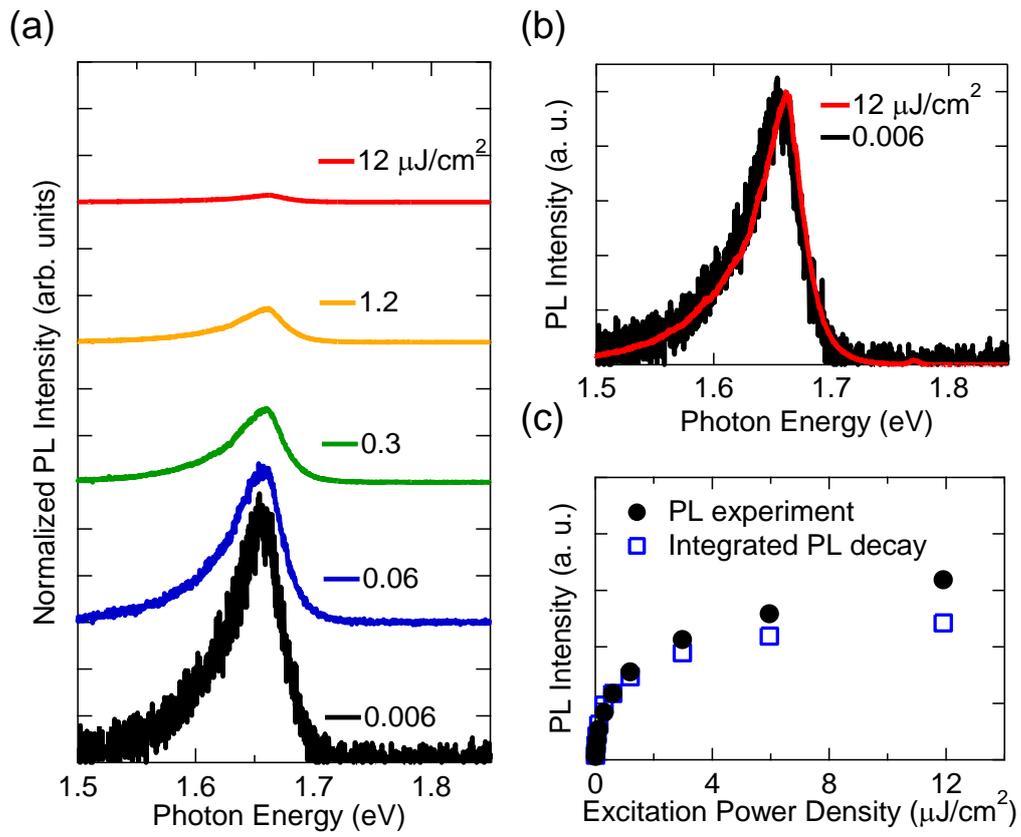

Fig. 2 S. Mouri et al.

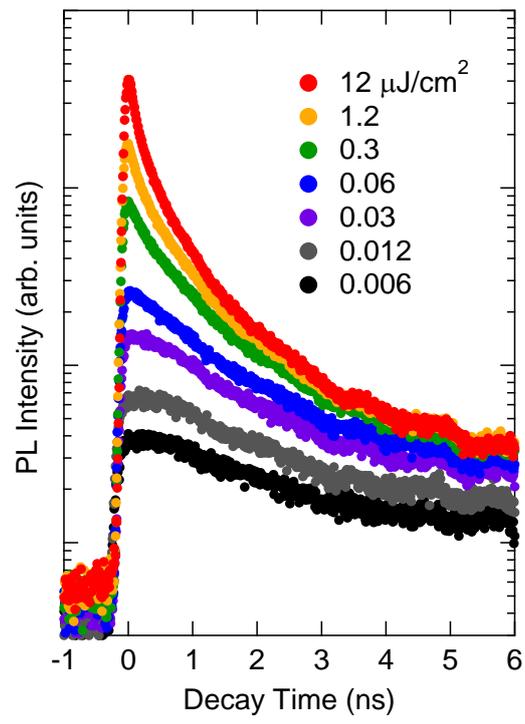

Fig.3 S. Mouri et al

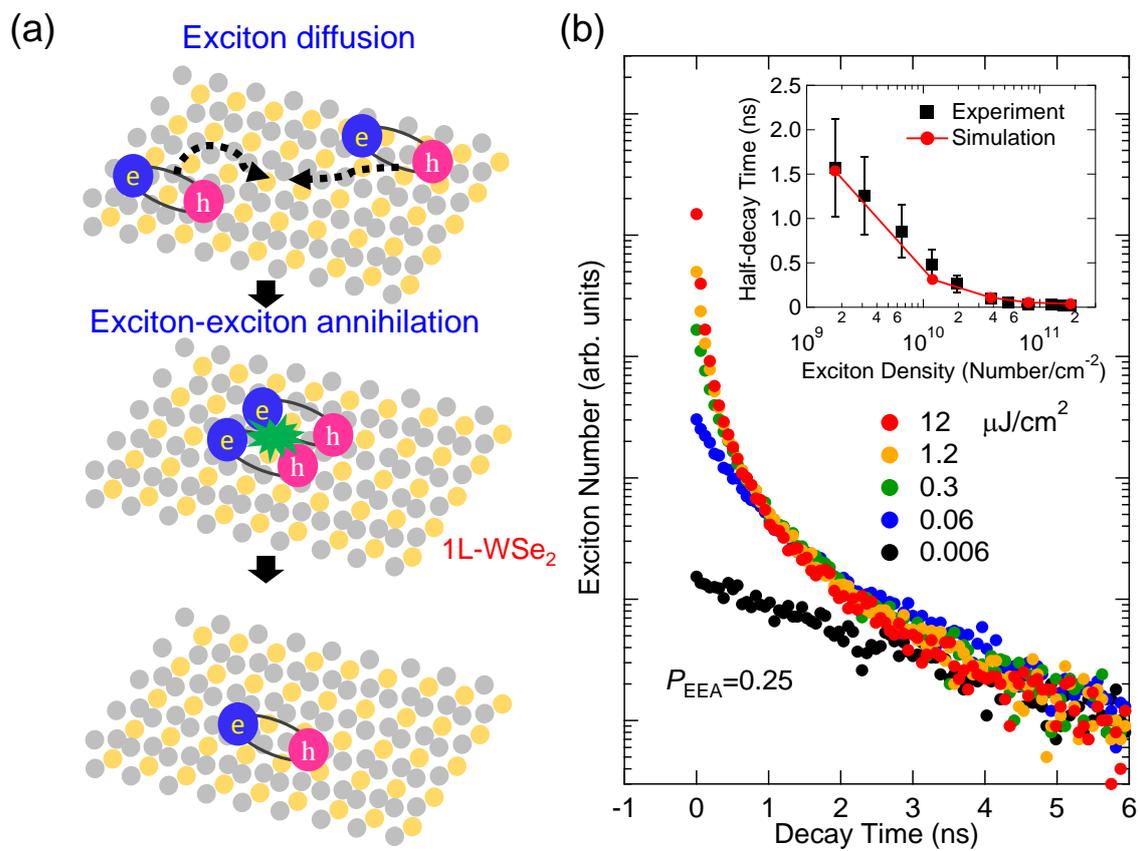

Fig.4 S. Mouri et al.